\documentclass[journal=jacsat,manuscript=article]{achemso}
\usepackage[version=3]{mhchem} 
\usepackage{xcolor}
\usepackage[T1]{fontenc}

\author{Rabbia Tahir}
\affiliation[Milano-Bicocca-PHYS] {L-NESS and Department of Physics, University of Milano-Bicocca, Milano, Italy}

\author{Paweł Wyborski}
\affiliation[DTU] {Department of Electrical and Photonics Engineering, Technical University of Denmark, Kgs. Lyngby, Denmark}

\author{Artur Tuktamyshev}
\affiliation[Milano-Bicocca] {L-NESS and Department of Materials Science, University of Milano-Bicocca, Milano, Italy}

\author{Stefano Vichi}
\affiliation[Milano-Bicocca] {L-NESS and Department of Materials Science, University of Milano-Bicocca, Milano, Italy}

\author{Richard~N\"otzel}
\affiliation[Milano-Bicocca] {L-NESS and Department of Materials Science, University of Milano-Bicocca, Milano, Italy}

\author{Battulga Munkhbat}
\affiliation[DTU] {Department of Electrical and Photonics Engineering, Technical University of Denmark, Kgs. Lyngby, Denmark}

\author{Stefano Sanguinetti}
\affiliation[Milano-Bicocca] {L-NESS and Department of Materials Science, University of Milano-Bicocca, Milano, Italy}
\email{stefano.sanguinetti@unimib.it}

\title[C-band quantum emitters by strain free QDs]
  {Strain-free, symmetrical, InGaAs quantum dots as single photon emitters in the telecom C-band}


\begin{document}


\begin{abstract}
 Non-classical photon sources made of semiconductor quantum dots (QDs) emitting in the telecommunication C-band are crucial components for low-loss, long-distance photonic quantum communication networks. Here we designed and fabricated strain--free In$_{0.7}$Ga$_{0.3}$As/In$_{0.7}$Al$_{0.3}$As QDs grown on GaAs(111)A substrates working as single-photon emitters in the 1550 nm window. The QDs were grown via local droplet etching method in a molecular beam epitaxy environment, employing a thin In$_{0.7}$Al$_{0.3}$As metamorphic buffer layer with the same lattice constant of the QD material, thus allowing for a completely strain--free self-assembly of the QDs. The QDs exhibit a C$_{3v}$ symmetry with a ground state emission in the 1400--1600 nm range. The exciton lifetimes of $\approx$ 1.3--1.9 ns and linewidths as low as $\approx$ 300 $\mu$eV show the good quality of the fabricated QDs. Second-order autocorrelation measurements under pulsed excitation confirmed the single-photon purity of the emitters, yielding a $g^{(2)}(0)$ value of $0.141 \pm 0.027$.
\end{abstract}


\section {Introduction}

Photonic quantum communication relies on the availability of system capable of generating and transmitting well-defined quantum states by means of single photons. These photons are regarded as a source of transporting quantum information to the specified quantum nodes for storage and processing \cite{huber2018semiconductor}. For the long-distance transmission via the existent fiber optics network, ideally photons should reside inside the telecommunication C-band, specifically 1530--1565 nm, to minimize the transmission losses \cite{michl2023strain}. Semiconductor quantum dots (QDs) have emerged as one of the prominent candidates for quantum emitters due to their properties including engineered discrete energy levels and the ability to generate single photons and entangled photon pairs on demand \cite{deutsch2023telecom, wyborski2025high, da2021gaas, tomm2021bright, zhai2020low}. 

Several studies demonstrated the fabrication of various QDs, such as In(Ga)As/GaAs \cite{semenova2008metamorphic,veretennikov2025single, scaparra2024broad, spitzer2024telecom, sittig2022thin} and In(Ga)As/InP \cite{sala2024self, smolka2021optical, joshi2023inp, ge2024polarized, birowosuto2012fast}, emitting in the telecom O- and C-bands. In(Ga)As QDs directly fabricated on the InP platform have the ability to naturally emit at telecom wavelengths. GaAs platform offers a high refractive index contrast of the GaAs-AlGaAs interface crucial for the construction of photonic structures as well as the monolithic integration of QDs with GaAs-based optoelectronic devices \cite{veretennikov2025single, wei2023monolithic}. Recently, GaSb material based system was also developed for the telecommunication system having high spectral purity \cite{hakkarainen2024telecom, masson2026engineering}. 

The fabrication of QDs can be accomplished through diverse growth strategies including the Stranski–Krastanov (SK) growth mode \cite{berdnikov2024near} as well as alternative approaches such as droplet epitaxy (DE) \cite{gurioli2019droplet, somaschini2010self, skiba2017universal, tuktamyshev2019temperature} and local droplet etching (LDE) \cite{deutsch2023telecom, deutsch2025local}. 
The latter self-assembly strategy, LDE creates QDs by etching nanoholes in molecular beam epitaxy (MBE) environment at relatively high temperature and then filling the holes with the appropriate material for quantum confinement \cite{heyn2015dynamics, tuktamyshev2024local, zocher2019droplet}. The two-step procedure enables for highly designable QDs in terms of composition, aspect ratio, and density \cite{covre2026low}. Elevated temperature can thereby enable the formation of QDs with less structural defects, hence improving their optical quality \cite{da2021gaas}. Its mechanism involves the deposition of group III atoms under low background pressure of group V material for the formation of droplets or specifically the etching of holes at high temperature after the nucleation of droplets.

Extending the emission of In(Ga)As QDs, grown on GaAs substrates, into C-band needs substrate strain engineering. This can be accomplished by employing thin compositionally graded metamorphic buffer layer (MMBL) as virtual substrates to assist the gradual strain relaxation \cite{scaparra2024broad, sittig2022thin}. However, its challenging to grow high quality thin MMBL on conventional GaAs(001) substrate due to the strain induced defects formation. This difficulty promotes the exploration of (111)--oriented surfaces at which complete strain relaxation of metamorphic buffer with thickness slightly exceeding 40 nm can be achieved due to the rapid plastic relaxation via the formation of misfit dislocations at the interface \cite{tuktamyshev2022flat, ohtake2020strain}. Additionally, (111)--oriented surfaces have inherent three fold rotational symmetry which is beneficial for the creation of symmetrical QDs \cite{basso2018high, mano2010self}. It leads to the reduction of the fine structure splitting (FSS) and enhances the photon indistinguishability of QDs and high entanglement fidelity \cite{huber2017highly}. For this purpose, previously InAs/InAlAs QDs were grown on GaAs(111)A substrates with an emission wavelength of around 1300 nm \cite{Tuktamyshev2021, Barbiero2022, wyborski2025high}. Extending the emission of these QDs further into C-band requires significantly larger QDs size which may introduce the defects due to excessive strain accumulation \cite{tuktamyshev2022strain}. 

Here we present strain--free InGaAs QDs with C-band single photon emission (i.e. 1550 nm) on InAlAs metamorphic buffer layer realized on vicinal GaAs(111)A substrate. In order to obtain low density, high optical quality QDs, the fabrication choice fall on the LDE technique as it allows for a large variety of materials compositions of the QDs and of the barrier, while maintaining the flexibility of droplet-driven epitaxy in terms of fabricated QDs size, shape and density.

\section {Results and Discussion}

\begin{figure}[h!]
\centering
\includegraphics[width=0.8\textwidth]{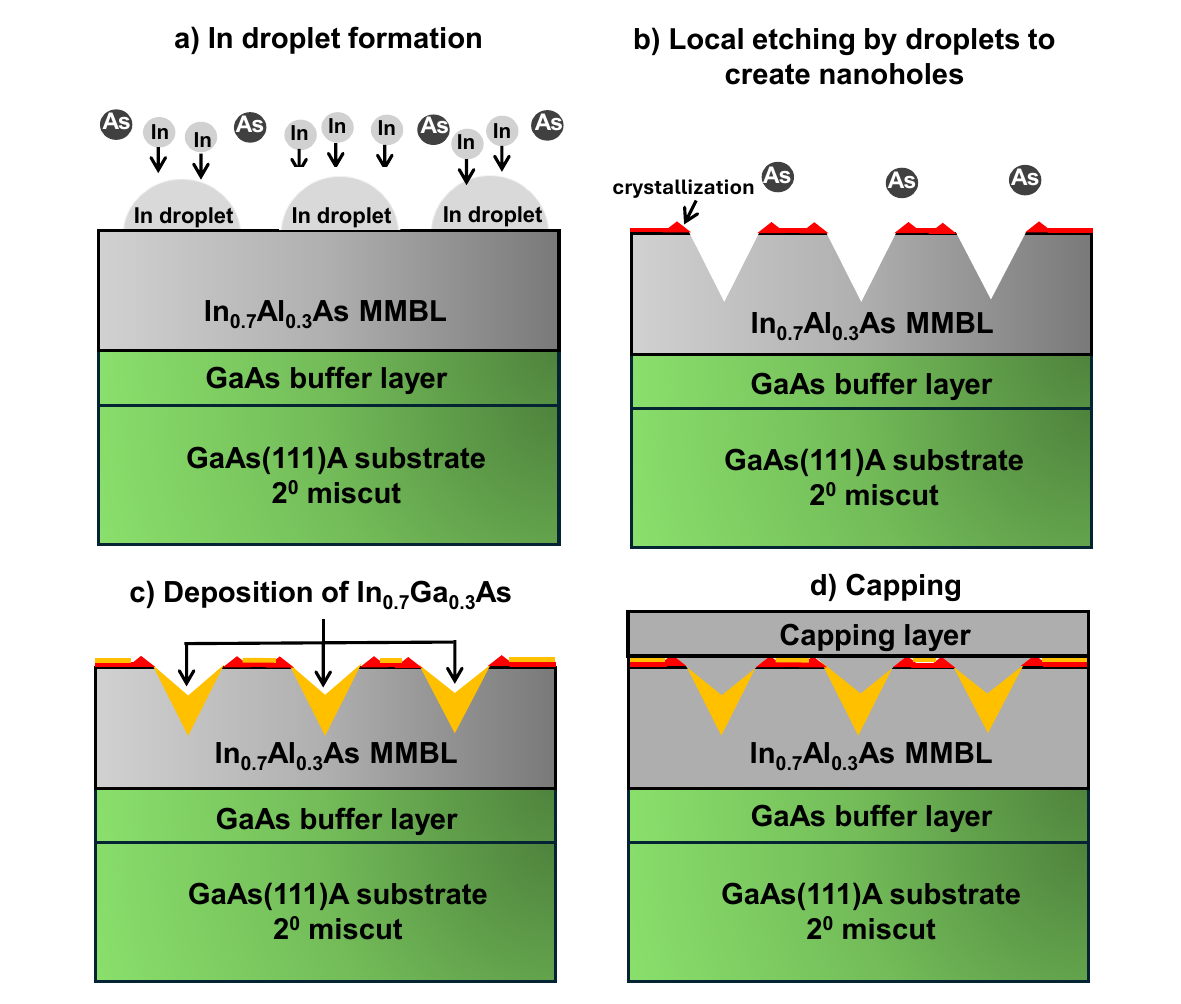}
\caption{\label{fig:1} Schematic diagram, summarizing steps for the InGaAs/InAlAs QDs growth by LDE on GaAs(111)A substrate with InAlAs MMBL such as (a) formation of droplets after depositing In, (b) local surface etching in As$_4$ background atmosphere by the indium droplets and creation of nanoholes, (c) filling the nanoholes with InGaAs to fabricate required QDs, (d) growth of a barrier capping layer.}
\end{figure} 

The LDE process is schematized in Figure \ref{fig:1}. The procedure starts with the deposition of metal droplets (Figure \ref{fig:1}a). Here the droplet size and density are controlled by the metal flux, the substrate temperature during the deposition, and the material amount \cite{heyn2015dynamics}. Subsequent annealing of the droplets in an As$_4$ atmosphere promotes substrate etching directly beneath the droplets (Figure \ref{fig:1}b). In this stage, a complex interplay between droplet size, arsenic flux, and substrate temperature determines the resulting nanohole shape, width, and depth \cite{heyn2011kinetic}. The nanoholes are then filled with the QDs material through preferential diffusion and accumulation inside the nanoholes \cite{da2021gaas}(Figure \ref{fig:1}c). The final QDs dimensions are dictated by the nanohole morphology and the volume of deposited material for the filling. To complete the structure, an upper barrier layer is grown as a cappig layer (Figure \ref{fig:1}d).

The electronic structure of the LDE-QDs is then governed by several parameters: i) the barrier composition, ii) the morphology of the nanohole, and iii) the composition and amount of the filling material. Notably, to remove the presence of strain, suppress defect formation and meet specific emission requirements, the InAlAs barrier and the InGaAs QDs share the same indium content. 

To optimize the InAlAs/InGaAs compositions and QDs size for the targeted C-band emission, the electronic properties were simulated by using the envelope function approximation (EFA) framework with an eight band k$\cdot$p approach \cite{auf2007multiscale}. The nanoholes geometry was modeled according to the typical shape obtained by the local droplet etching on (111) substrates, a triangular inverted pyramid \cite{tuktamyshev2024local}, which is dictated by the surface symmetry and the presence of low-etch-rate $\{111\}$ facets that act as etching limits. Based on our simulations, the optimal compositions for the barrier and QDs were determined to be In$_{0.7}$Al$_{0.3}$As and In$_{0.7}$Ga$_{0.3}$As, respectively, with a target QDs height of approximately 6~nm and an aspect ratio (QDs height/QDs width) in the range of 0.10--0.15.

Since the actual size and shape of droplet-etched nanoholes strongly affect the electronic properties of QDs and depend on growth parameters such as droplet material, droplet size and annealing conditions, we conducted a comprehensive analysis of the achievable morphology of QDs by LDE in In$_{0.7}$Al$_{0.3}$As MMBL by changing these parameters, as reported in the Supporting Information (Supporting Note S1).  In order to characterize the effects of droplet etching and nanohole filling, two samples were grown by stopping the procedure after the annealing of the droplets in As$_{4}$ atmosphere (sample A) and after the deposition of the InGaAs layer (sample B).

\begin{figure}[h!]
\centering
\includegraphics[width=1\textwidth]{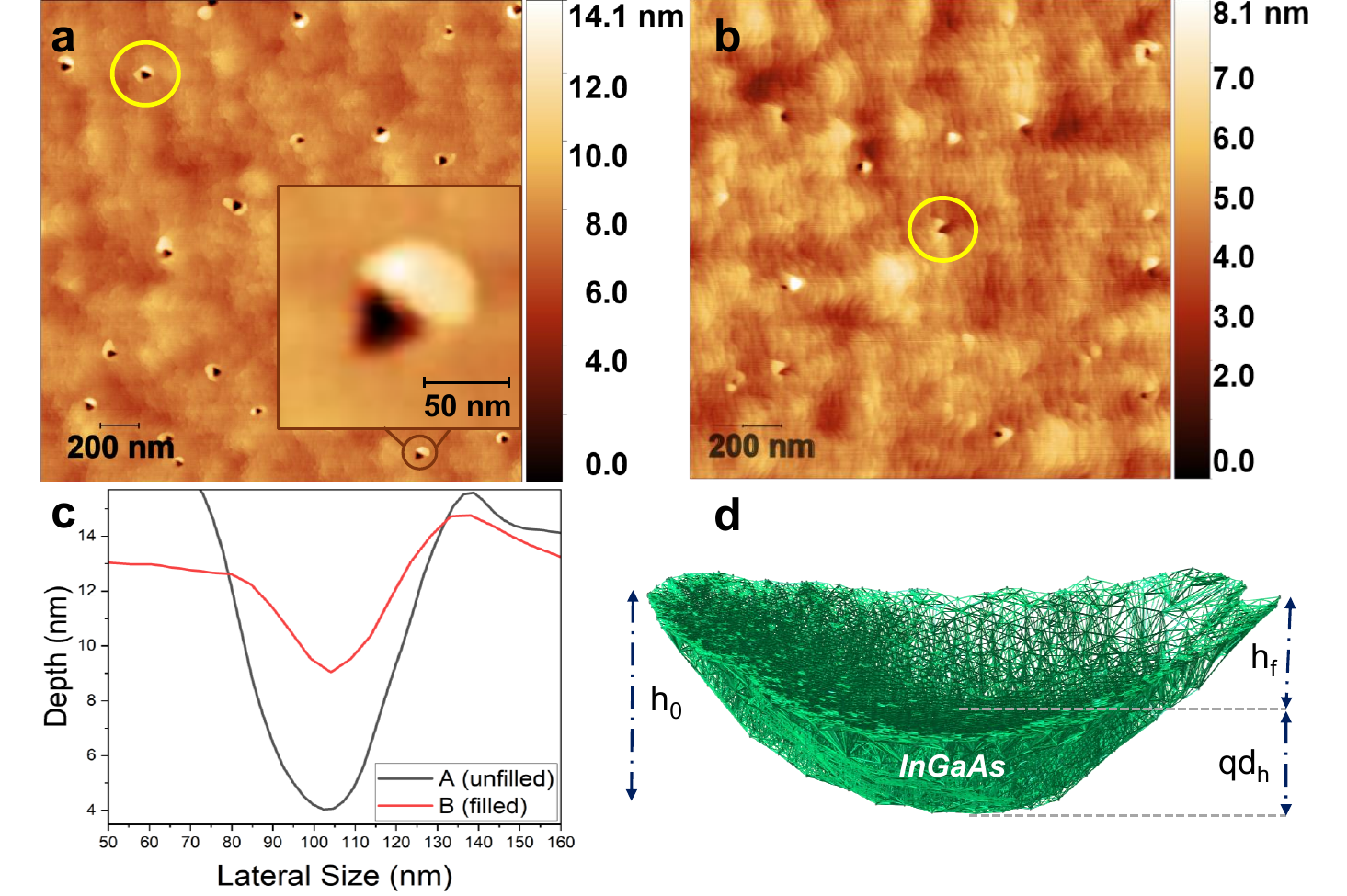}
\caption{\label{fig:3} (a) 2.5$\times$2.5 $\mu$$m^2$ AFM topography image of sample A. The  inset reports the magnified image of 6.8 nm deep and 54.3 nm wide individual nanohole. (b) 2.5$\times$2.5 $\mu$$m^2$ AFM topography image of sample B.  (c) Height cross-section profiles of both unfilled and filled nanoholes from sample A and B. The two nanoholes in (a) and (b) are highlighted by yellow circles. (d) 3D mesh of the AFM-derived sloped geometry of grown QDs with an effective height of 5.13 nm inside the InAlAs barrier, i.e. reconstructed from the difference between the  extracted profile of the nanohole (highlighted by yellow circle in (a), which has the mean height of 8.45 nm and mean lateral size of 51.0 nm) used in the simulation.}
\end{figure}

Atomic Force Microscopy (AFM) images of the final set of fabricated nanoholes and QDs, that best matches the design configuration described in the experimental part, are shown in Figure \ref{fig:3} (a, b). In these images, the dark regions correspond to the nanoholes, while the lighter surrounding areas represent the material displaced during the droplet etching process that crystallizes close to the droplets \cite{heyn2015dynamics}.

Each nanohole in Figure \ref{fig:3}(a) (sample A) exhibits well defined and symmetric morphology characterized by a triangular pyramid shape having average depth and lateral size as $8.45\pm2.2$ nm and $51.0\pm8.7$ nm, respectively. As expected, the dimensions of nanoholes i.e. width and depth were dependent on the droplet volume and the annealing conditions \cite{heyn2015dynamics}. The nanohole density was calculated as $3.84 \pm 0.78 \times 10^8$/cm$^{2}$. 
Figure \ref{fig:3}(b) reports the morphology of overgrown surface with 2~nm of InGaAs (sample B). It is possible to observe the presence, even in the overgrown sample, of shallow nanopits with an average height and lateral size of $3.32\pm1.0$ and $57.3\pm10.1$, respectively, and the density of $1.72 \pm 0.26 \times 10^8$/cm$^{2}$. We attribute the reduction of the density and of the average depth of the nanoholes to the successful filling of etched nanoholes by the deposited InGaAs. While the shallower nanoholes are completely filled, the larger becomes less deep due to the accumulation of material at the bottom. The actual height of the QDs ($qd_h$) can be estimated by taking the difference between the height obtained for unfilled sample A ($h_0$) and height obtained for filled sample B ($h_f$) from AFM measurements. The fabricated QDs showed a sloped geometry (see Figure \ref{fig:3}(c)), consistent with the typical shape of strain free LDE-QDs \cite{heyn2022dot}.

The measured AFM topography of the nanoholes before and after InGaAs deposition, was then utilized as a direct input to simulate the emission properties of the fabricated QDs (see Figure \ref{fig:3}(d)).  The final structure (similar as already reported shape \cite{heyn2022dot}) was then simulated, resulting in a calculated transition energy of 0.81~eV that corresponds to the emission of 1530~nm for fabricated QDs (see more details in Supporting Note S2). It is worth mentioning that the In$_{0.7}$Al$_{0.3}$As barrier transition is expected at 1159~eV (see Supporting Figure S3). 
 
The optical properties of the fabricated QDs were investigated by photoluminescence (PL) spectroscopy at 4 K (optical setup description included in Methods and Supporting Note S3). The low density of QDs allows direct single-QD characterization by limiting collection to a focused laser for excitation.

\begin{figure}
\centering
\includegraphics[width=0.9\textwidth]{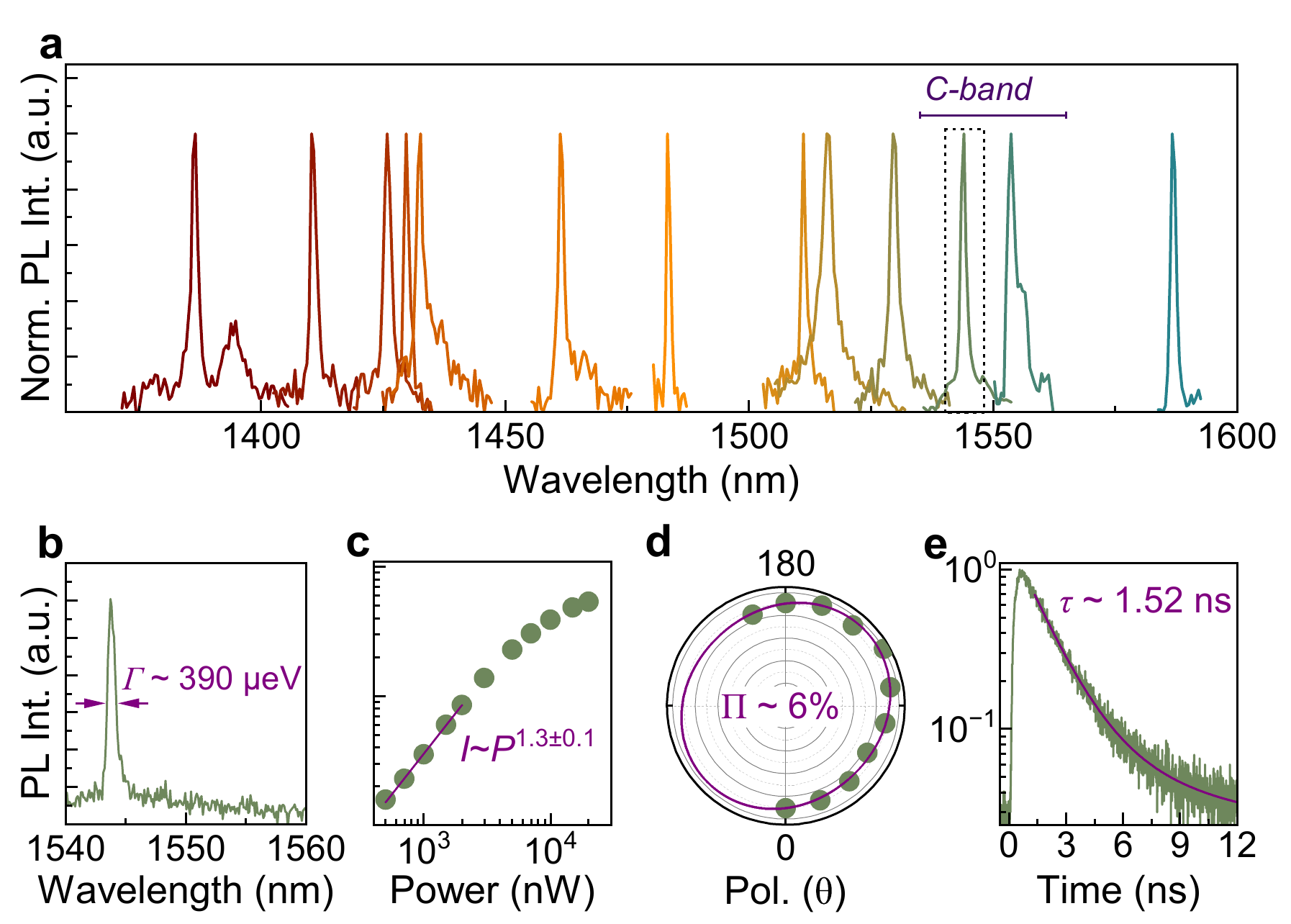}
\caption{\label{fig:5}(a) $\mu$-PL spectra of several QDs with sharp emission peaks centered at around 1380--1600 nm. (b) $\mu$-PL spectra of marked QD in (a) (within C-band range), including fitted line broadening value. (c) Power-dependent QD PL intensity fitted by a power law. (d) Emission intensity plot as a function of polarization, including DOLP fitting. (e) Time-resolved PL histogram of QD emission, with the solid lines depicting a monoexponential fit to the experimental data.}
\end{figure}

Figure~\ref{fig:5}(a) depicts the emission lines obtained from several individual QDs. The spectra exhibit multiple lines centered in the range of about 1380--1600 nm, attributed to the ground-state transition of the grown QDs. These intense peaks agree well with theoretical simulations that predicted the emission wavelength of the fabricated QDs (as shown in Figure~\ref{fig:3}(a)) in the C-band range, and thus validate the accuracy of the geometrical modeling performed before and after the growth. The macro-PL spectra, shown in the Supporting Figure~S4, support these observations, with ensemble QD emission observed in the 1400--1600 nm range.

Figure \ref{fig:5}(b-e) shows more detailed characterization results for the representative QD within the C-band, emitting at around 1544 nm (marked with a dashed curve in Figure~\ref{fig:5}(a)). Additionally, Supporting Figure~S5 includes a similar analysis of two other QDs emitting at different wavelengths, 1455 and 1586 nm, respectively. In particular, we present here an analysis of high-resolution $\mu$-PL spectra, a power-dependent PL intensity curve, in-plane polarization-dependent PL signals, and time-resolved PL histograms to verify the quality and properties of the grown QDs. Additional comparisons of QDs emitting across different spectral ranges provide deeper insights into how size variations affect their intrinsic properties. Figure~\ref{fig:5}(b) presents the QD spectrum showing Gauss-profile-based fitted line broadening of around 390 $\mu$eV with similar values observed for other QDs (300 and 550 $\mu$eV as shown in the Supporting Figure~S5). Obtained linewidth values are slightly broader than already reported values for C-band InAs/GaAs QDs (i.e., 270 $\mu$eV) grown on a metamorphic layer with a Bragg reflector \cite{wronski2021metamorphic}. They are also broader than linewidth emission of about 30--120 $\mu$eV range for InAs/InP QDs emitting in telecom C-band \cite{smolka2021optical}. The observed broader linewidth values are likely due to spectral diffusion, charge fluctuations, or defects in the barrier region around QDs \cite{abbarchi2008spectral}. This finding is consistent with the reason that larger QDs exhibit broader linewidths, as the enhanced spatial extent of the exciton in larger QDs can increase the sensitivity to local charge fluctuations, leading to faster dephasing and broader linewidth \cite{zhai2020low, Wyborski2023Impact}. 

PL intensity of the single QDs follows the power law as $I \propto P^\alpha$ after exciting with optical power $P$ \cite{paur2019electroluminescence}. The exponent values $\alpha$ of unity and 2 are associated with excitonic and biexcitonic recombinations, respectively, while the deviation between these values corresponds to mixed states or additional contributions, such as carrier dynamics around the QDs. However, in power-dependent PL spectra as presented in Figure~\ref{fig:5}(c), the $\alpha$ values are extracted as $1.3\pm0.1$, affirming that the emission originates not only from the single-exciton radiative transition but also from mixed states \cite{hu2020photoluminescence, li2018revealing}. Figure \ref{fig:5}(d) shows the emission intensity as a function of polarization with fitted degree of linear polarization (DOLP, symbolically abbreviated as $\Pi$) that defines the polarization asymmetry of QDs emission, which can be induced by the mixture of heavy and light holes states \cite{abbarchi2021polarization, Henini1998}. The obtained DOLP value of about 6\%, as well as 13\% and 22\% for other QDs (see Supporting Figure~S5), showed rather limited polarization anisotropies with only slightly enhanced emission in one polarization direction. In-plane asymmetry of QD shape is one of the origins for these anisotropies by modifying the mixing of valence band states \cite{wyborski2022electronic}. It can also be due to the accumulation of material (clearly visible in AFM images in Figure \ref{fig:3}) around the holes, which may correspond to the asymmetries. Nevertheless, observed values below 30\% suggest rather limited shape anisotropies for the investigated QDs and likely indicate possible realization of the low FSS QDs.

Moreover, a time-resolved PL histogram for the investigated QD was also obtained, as shown in Figure~\ref{fig:5}(e). The monoexponential decay fitting yields a lifetime of about 1.52 ns (with similar values of 1.73 ns and 1.91 ns for other QDs shown in Supporting Figure~S5). The lifetimes are in good agreement with already reported data for the QDs emitting in the telecommunication range \cite{chen2016telecommunication, phillips2024purcell, wronski2021metamorphic, anderson2021coherence, smolka2021optical}. The slight variation in the lifetimes of QDs emitting at different wavelengths might be due to differences in QD size and composition. 

Second-order autocorrelation $g^{(2)}(0)$ measurements were performed under pulsed excitation to affirm the single photon behavior of the grown QDs \cite{ge2024polarized}.

\begin{figure}
\centering
\includegraphics[width=1\textwidth]{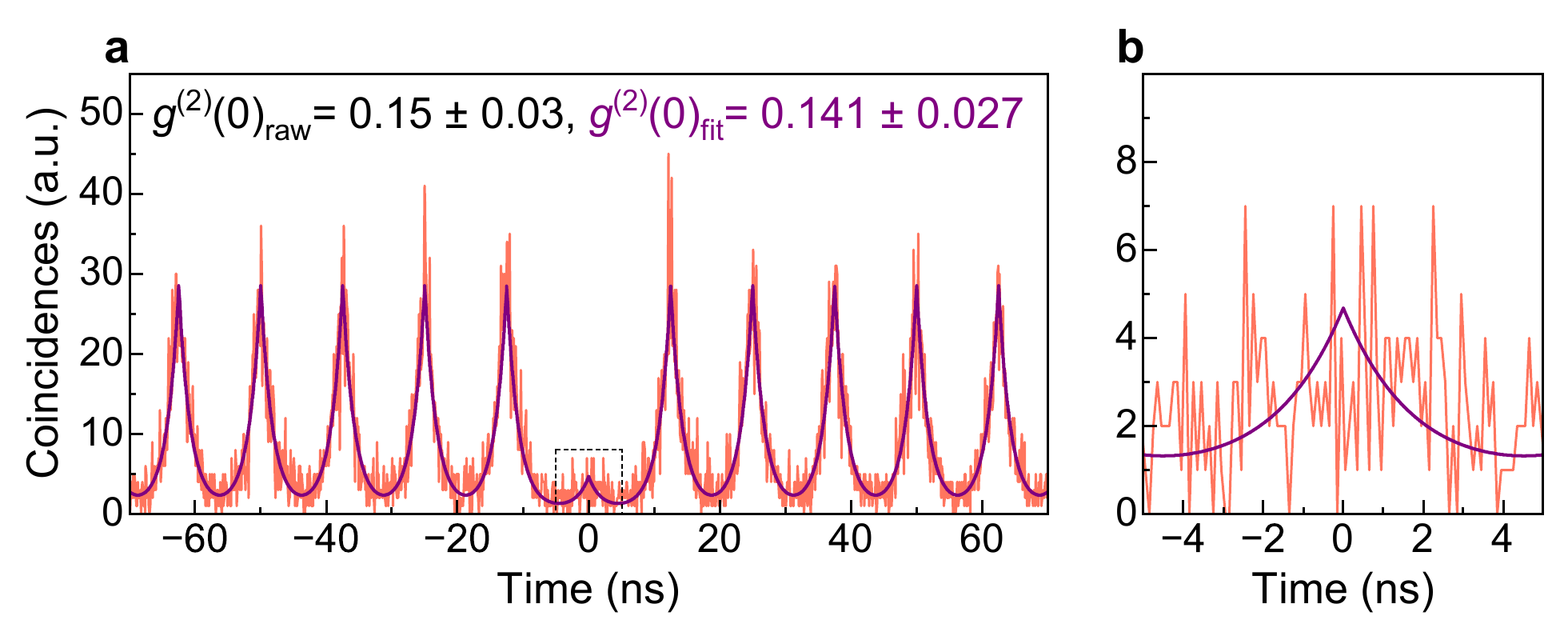}
\caption{\label{fig:7} (a) Autocorrelation coincidence plot. (b) Second order correlation function $g^2(0)$ for CX emission, the solid purple line represents the direct fit for the result.}
\end{figure}

Figure~\ref{fig:7}(a, b) depicts the autocorrelation measurements of one of the brightest grown QD and the as-measured $g^{(2)}(0)$ value as $0.15\pm0.03$, confirming the emission of single photons. Slightly improved values of $0.141\pm0.027$ under pulsed excitation was obtained after fitting the autocorrelation data to reduce the noise contribution. These non-zero values with some errors are likely to have been originated from background emission \cite{le2017temperature}. It can also result from using higher excitation power to obtain sufficient coincidence counts, which can increase background radiation. The observed $g^{(2)}(0)$ values are consistent with the recent development of InAs/InGaAs QDs that have a $g^{(2)}(0)$ value of 0.14 for C-band \cite{veretennikov2025single}. 

In conclusion, we have successfully demonstrated single-photon emitters operating in the telecommunication C-band, based on In$_{0.7}$Al$_{0.3}$As/In$_{0.7}$Ga$_{0.3}$As QDs grown on GaAs(111)A substrates. The QDs were realized using the LDE method, which enables precise engineering of the emitter electronic properties due to LDE inherent flexibility of QD composition and morphology. By tuning growth parameters, specifically the substratetemperature during the deposition, the annealing temperature, and material fluxes, we achieved a fine control over nanohole formation and subsequent QD geometries. Theoretical modeling was the basis for the QD design to identify the point in the parameter space (QD and barrier composition, nanohole shape and size) that could lead us to the right QD emission. Subsequent calculations based on the actual morphology of the fabricated nanostructures yielded simulated emission spectra in excellent agreement with experimental observations. The QDs exhibited linewidth ranging from 300 to 550 $\mu$eV and exciton lifetimes between 1.52 and 1.91 ns. Furthermore, single-photon emission was confirmed through second-order autocorrelation measurements, reaching a $g^{(2)}(0)=0.141 \pm 0.027$. 

The high degree of freedom made possible by LDE for the design QD shape and composition provides a unique predictive framework for fabricating high-quality C-band InGaAs/InAlAs QDs, representing a significant step forward for long-distance telecommunication. Still, the relatively broad linewidth, the long lifetimes and $g^{(2)}(0)$ values suggest room for optimization. Future improvements could involve stabilizing the local charge environment and mitigating trap-related charge transfer by embedding the QDs within a p-i-n diode structure, thereby approaching the performance of state-of-the-art telecom photon emitters. By making use of the compatibility of the presented InGaAs LDE-QDs with photonic structures, we expect that their radiative lifetimes could be reduced to few hundreds of picoseconds via the Purcell enhancement of the radiative coupling \cite{phillips_purcell-enhanced_2024}. This would allow  for quantum light sources with better photon indistinguishability and that can be driven at rates well above 1 GHz.

\section {Experimentals}

\subsection{Growth}
Samples were grown on GaAs(111)A substrate with a miscut of $2^\circ$ towards $(\overline{1}\overline{1}2)$ direction via LDE method in Molecular Beam Epitaxy chamber. In droplets were used for creating the nanoholes (see Supporting Note S1). Vicinal GaAs(111) substrate was utilized because it promotes step-flow growth mode thus permitting the growth of flat surface with the growth rate equivalent to the standard one on GaAs(001) \cite{tuktamyshev2022flat, herzog2012optimization}.

For producing the confining barrier material for the QDs, In$_{0.7}$Al$_{0.3}$As MMBL with optimized composition (see Supporting Note S1) was deposited at the substrate temperature $T_s$ of 450~$^\circ$C on GaAs with the growth rate of 0.5 monolayers per second (ML/s) (where 1~ML corresponds to the 6.23$\times$10$^{14}$ atoms/cm$^{2}$ density of the GaAs(001) unreconstracted surface). The thickness of MMBL is 150~nm i.e. much greater than 40~nm which is required for the complete strain relaxation of the system \cite{tuktamyshev2022flat}. Before indium droplet nucleation, As valve and As shutter were closed to minimize As background pressure. Then 1.05~ML of In was deposited with a rate of 0.35~ML/s with the background As$_4$ pressure of 1$\times 10^{-7}$ torr and $T_s$ of 500~$^\circ$C. The etching process step was performed under the same As$_4$ pressure and substrate temperature for 1 min and then As valve and As shutter were open for 1 min resulting in As$_4$ beam equivalent pressure (BEP) of 1.1$\times 10^{-5}$ torr. A subsequent deposition of 2-nm thick In$_{0.7}$Ga$_{0.3}$As layer was performed with a growth rate of 0.5~ML/s and $T_s$ of 500~$^\circ$C followed by 2 min annealing at the same conditions to fill the nanoholes and fabricate the QDs. Subsequently, the deposition of 100~nm In$_{0.7}$Al$_{0.3}$As barrier layer, followed by 3~nm of In$_{0.7}$Ga$_{0.3}$As capping at $T_s$ of 450~$^\circ$C complete the structure. The schematic procedure for the creation of QDs is depicted in the Figure \ref{fig:1}. In order to characterize the effects of droplet etching and nanohole filling, two samples were grown by stopping the procedure after the annealing of the droplet in As$_{4}$ atmosphere (sample A) and after the deposition of the InGaAs layer (sample B).

\subsection{Morphological Characterization}
The morphology of the samples was investigated using the Atomic Force Microscope (AFM) in standard tapping mode at room temperature with sharp silicon tips with a lateral resolution of about 2~nm. The lateral size of the nanoholes was measured as the altitude of the triangular base parallel to the x-scan direction. The difference between the minimum point of the nanohole and the surface level determines the nanohole's depth.

\subsection{Theoretical Calculations}

The electronic properties were simulated by using the envelope function approximation (EFA) framework with an eight band k$\cdot$p approach \cite{auf2007multiscale} using the actual QD shape as input. The simulation parameters, including band gaps and effective masses, are reported in Supporting Note S2, tables TS2 and TS3. 

\subsection{Optical Characterization}
The measurement of optical properties of QDs were performed via a spectrometer with 0.328 m focal-length monochromator (Kymera 328i, gratings: 150, 600 lines/mm, Andor, Oxford Instruments) and a deep thermo-electrically-cooled (In, Ga)As linear array detector (iDus, Andor, Oxford Instruments) was used for PL spectra recording and spectral filtering. The polarization-resolved PL characterizations were obtained with a pair of a half-wave plate and a linear polarizer. Time-resolved PL characterization and single photon statistics measurements using a Hanbury Brown and Twiss configuration (based on a 50:50 fiber beam-splitter) were performed using time-correlated single-photon-counting mode based on a single photon counting module (Time Tagger Ultra, Swabian Instruments) together with superconducting nanowire single-photon detectors (ID281 SNSPD, ID Quantique).

\begin{acknowledgement}

R.T, S.V., A.T., R.N. and S.S. acknowledge support from the PNRR MUR project ‘‘National Quantum Science and Technology Institute’’ - NQSTI (PE0000023). P.W. and B.M. acknowledge support from the European Research Council (ERC-StG ``TuneTMD'', grant no. 101076437) and Villum Fonden (project no. VIL53033), as well as the European Research Council (ERC-CoG ``Unity'', grant no. 865230) and the TICRA foundation, the Innovation Fund Denmark (QLIGHT, no. 4356-00002A).

\end{acknowledgement}

\begin{suppinfo}

Detail on optimization of QDs design, construction and simulation of AFM-derived geometries, Additional information on ensemble and single QDs optical characterization.

\end{suppinfo}

\bibliography{references}

\end{document}